\documentstyle[11pt,aaspp4,tighten]{article}

\begin{document}

\font\twelvei = cmmi10 scaled\magstep1 
       \font\teni = cmmi10 \font\seveni = cmmi7
\font\mbf = cmmib10 scaled\magstep1
       \font\mbfs = cmmib10 \font\mbfss = cmmib10 scaled 833
\font\msybf = cmbsy10 scaled\magstep1
       \font\msybfs = cmbsy10 \font\msybfss = cmbsy10 scaled 833
\textfont1 = \twelvei
       \scriptfont1 = \twelvei \scriptscriptfont1 = \teni
       \def\mit{\fam1 }
\textfont9 = \mbf
       \scriptfont9 = \mbfs \scriptscriptfont9 = \mbfss
       \def\bmit{\fam9 }
\textfont10 = \msybf
       \scriptfont10 = \msybfs \scriptscriptfont10 = \msybfss
       \def\bmsy{\fam10 }

\def\etal{{\it et al.~}}
\def\eg{{\it e.g.}}
\def\ie{{\it i.e.}}
\def\lsim{\raise0.3ex\hbox{$<$}\kern-0.75em{\lower0.65ex\hbox{$\sim$}}} 
\def\gsim{\raise0.3ex\hbox{$>$}\kern-0.75em{\lower0.65ex\hbox{$\sim$}}} 
 
\title{Solutions of Two Dimensional
Viscous Accretion and Winds In Kerr Black Hole Geometry\footnote[1]
   {Submitted to the Astrophysical Journal}}
 
\author{Sandip K. Chakrabarti}
\affil{Tata Institute of Fundamental Research, Bombay 400005, India;\\
   e-mail: chakraba@tifrc2.tifr.res.in}

\vskip 1cm
\begin{abstract}

We extend our previous studies of shock waves and shock-free solutions 
in thin accretion and winds in pseudo-Newtonian geometry to the case 
when the flow is {\it two-dimensional} and around a {\it Kerr black hole}.
We present equations for fully general relativistic viscous transonic flows and
classify the parameter space according to whether or not 
shocks form in an inviscid flow. We discuss the 
behaviors of shear, angular momentum distribution, heating and cooling
in viscous flows. We obtain a very significant
result: we find that in weak viscosity limit the presence of the standing shock waves
is more generic in the sense that flows away from the equatorial
plane can produce shock waves in a wider range of parameter space. Similar
conclusion also holds when the angular momentum of the black hole
is increased. Generally, our conclusions regarding the shape of the
shock waves are found to agree with results of the existing  numerical 
simulations of the two dimensional accretion in Schwarzschild geometry.
In a strong viscosity limit, the shocks may be located farther out or
disappear completely as in the pseudo-Newtonian geometry.

\end{abstract}


\noindent  Appearing in Astrophysical Journal On Nov. 1st, 1996

\clearpage
 
\section{INTRODUCTION}

The recognition that the accretion process onto compact objects
are inherently  three-dimansional is as old as the
subject of accretion itself. The pioneering work of Bondi (1952), however,
contained the solution for spherically symmetric inflow,
thereby reducing the problem in one dimension. Subsequently,
Lynden-Bell (1978) considered the flow to have a constant angular
momentum but without any radial velocity. He considered the
axisymmetric case, so that the problem is two dimensional,
but still integrable because of the choice of negligible radial
motion. This study was later extended to understand 
thick accretion flow structures in pseudo-Newtonian
geometry (e.g., Paczy\'nski \& Wiita, 1980) as well as in full
general relativity (Chakrabarti 1985, hereafter C85 and references therein)
using more general power law distribution of angular momentum
which mimicked a viscous rotating flow.

Nonetheless, major progress on the theoretical front has been done 
mostly for accretion flows which are thin and confined close to the equatorial
plane. The study of optically thick viscous Keplerian disks (Shakura \& Sunyaev, 1973;
Novikov \& Thorne, 1973), and viscous transonic disks (Paczy\'nski \& Bisnovatyi-
Kogan, 1981; Chakrabarti, 1990, hereafter C90; 
for a general discussion, see, Chakrabarti 1996a, 1996b, hereafter C96b,
and references therein) are some of the examples. 
Secondly, most of these studies are done using the so-called 
pseudo-Newtonian potential of Paczy\'nski \& Wiita (1980) which
mimics the external geometry of Schwarzschild black hole quite
well, although the computational error could be significant very close to the
horizon since the inner boundary condition is not exactly satisfied.
(see, Yang and Kafatos, 1995, for a discussion on isothermal disks in Schwarzschild geometry) 
Of interest to us in the context of the
present paper is a significant observation from the study of the 
most general viscous transonic flow that in a large range of
parameter space, spanned by the two free parameters, namely the
specific energy and angular momentum, the flow
develops standing shock waves close to a black hole (C90, C96b).
Questions naturally arise: Do these shocks continue to be present
when a full general relativistic study is made? If so, do the
general dependences of shock behavior on flow parameters remain
similar as in the pseudo-Newtonian treatment? Would these
shock waves also form {\it away} from the equatorial plane when the
flow is two dimensional? How would
the shock locations change as one goes from a non-rotating black hole
to a rotating black hole? In the present paper, we shall answer
these questions. Recent works (Chakrabarti \& Wiita, 1992; Chakrabarti \& 
Titarchuk, 1995) indicate that shock waves could be a natural sites of the hot
radiation in a black hole accretion, and that the observed quasi-periodic
oscillations could be due to some type of resonance oscillation
in the post-shock flow (Molteni, Sponholz \& Chakrabarti, 1996) or
due to dynamical oscillation of the sub-Keplerian flow (Ryu, Chakrabarti \& 
Molteni, 1996). These make it all the more important to understand the behavior of 
shock waves in accretion and wind flows close to a black hole.
Of course, a large region of the parameter space do not form
shocks at all. We therefore classify the entire parameter space for a 
conical wedge shaped flow to show which 
regions form shock waves, and which regions do not. Another
topic of importance  which did not fetch adequate attention in the 
literature so far, (despite being very relevant for the self-consistent treatment
of a viscous transonic flow) is: how should one treat the inner edge of the flow 
near the horizon? Traditionally, shear is treated to be zero
at the inner edge (at $r=r_{ms}$, the marginally stable orbit
for a Shakura-Sunyaev [1973] disk) or on the horizon for
pseudo-Newtonian viscous transonic flows (C90, C96b). Does shear indeed vanish 
on the horizon? If not, what effect, if any, would it have on the
angular momentum distribution, or heating processes? 
We plan to answer these questions as well. The motivation
is to build a general framework on which viscous transonic flows may be 
studied in future with full consistency when additional heating and cooling
effects are introduced.

Numerical simulations of fully two dimensional accretion
around a Schwarzschild black hole have already provided 
partial answers to some of the questions  posed above.
Hawley, Smarr \& Wilson (1984, 1985, hereafter HSW84, HSW85;
Hawley, 1984, hereafter H84) showed that quasi-spherical rotating flows 
develop traveling accretion shock waves whose shapes generally follow
the funnel walls around the vertical axis. More recent works of 
Molteni, Lanzafame \& Chakrabarti, 1994, hereafter MLC94; Ryu et al, 1995;
Molteni, Ryu \& Chakrabarti, 1996, hereafter MRC96) are examples of 
numerical simulations which produced {\it standing shock} waves in thin as well
as thick accretion flows. The importance of these latter works is
that the codes have been tested quite adequately against
the corresponding analytical transonic solutions (which 
were non-existent during the simulations of HSW84, HSW85)
for thin flows, and therefore the existence of standing 
shocks in thicker flows which they reveal may be trusted.

In the following sections, we shall try to solve the equations
of viscous transonic flows in an axisymmetric, thick flow 
from a purely analytical point of view. To achieve 
our goal we make simplifying approximations.
We also use fully general relativistic equations
for the flow. Accreting matter in this case automatically fulfills
the inner boundary condition on the horizon. 
We obtain solutions with and without shocks, both for
accretion and winds.
We arrive at several important conclusions regarding shock properties
in a thick flow:
(a) the region of flow parameters which enables shock formation is much bigger
in a rotating black hole. However, the general behavior 
of the parameter space itself is similar to what is obtained in 
in pseudo-Newtonian geometries (Chakrabarti 1989,
hereafter C89; C90; Sponholz \& Molteni, 1994) 
and (b) We find that the shock waves form away from the
equatorial plane more easily because of weaker gravity.
Since these shocks are centrifugally supported, a knowledge of the angular
momentum distribution is important. We therefore also explore the nature of the 
shear stress which governs the transport of angular momentum. Viscosity
not only transports angular momentum, it heats up the gas. Assuming the
viscosity is due to ions only we compute the heating rate of the
flow both in the subsonic and the supersonic branches. The cooling is governed
by the density and temperature of the flow. We present the variation of
the bremsstrahlung cooling rate along the flow as well, so that we may
find out the relative importance of heating and cooling effects which
might affect the flow properties.

The organization of the paper is as
follows: In Section 2, we present the basic mathematical equations
for the study of viscous, transonic flows in a black hole geometry. In Section 3,
we present a `brute force' method to find the general behavior
of the global surfaces, such as the funnel wall, centrifugal barrier,  
and the shock surfaces. In Section 4, we present the fully self-consistent
solution of the steady flow which may or may not
include shock waves. We then extend the result to a thicker flow. 
In Section 5, we summarize our work and make concluding remarks.

\section{MODEL EQUATIONS}

In what follows, we choose units of velocity, distance,
and time to be $c$, $GM_{bh}/c^2$ and $GM_{bh}/c^3$ respectively, where $G$
and $M_{bh}$ are the gravitational constant and the mass of the
black hole respectively. The matter distribution is assumed to be described by the
stress-energy tensor $T_{\mu \nu}$ which for a perfect fluid takes the form,
$$
T_{\mu \nu} = \rho u_\mu u_\nu + p (g_{\mu \nu} + u_\mu u_\nu ) ,
\eqno {(1)}
$$
which satisfies the equation of motion,
$$
\nabla^\mu T_{\mu\nu} = 0 ,
\eqno{(2)}
$$
Here, $p$ is the isotropic thermal pressure and $\rho=\rho_0(1+\pi)$ 
is the mass density, $\pi$ being the internal energy.
The four velocity components $u_\mu$ satisfy the normalization  relation,
$$
u_\mu u^\mu= -1
\eqno{(3)}
$$
where $\mu=0,1,2,3$. We assume the metric around a Kerr black hole in Boyer-Lindquist
coordinate (e.g., NT73),
$$
ds^2 = g_{\mu\nu}dx^\mu dx^\nu= -\frac{r^2 \Delta}{A} dt^2 + \frac{A}{r^2}
(d\phi-\omega dt)^2 +\frac{r^2}{\Delta} dr^2 + dz^2
\eqno{(4)}
$$
Where,
$$
A= r^4 + r^2 a^2 + 2 r a^2
$$
$$
\Delta= r^2 - 2 r + a^2
$$
$$
\omega = \frac{2 a r}{A}
$$
Here,  $g_{\mu\nu}$ is the metric coefficient and $u_\mu$ is the four velocity components:
$$
u_t= \left[\frac{\Delta}{(1-V^2)(1-\Omega l)(g_{\phi\phi}+l g_{t\phi})}\right]^{1/2}
$$
and
$$
u_\phi=-l u_t
$$
where, the angular velocity is
$$
\Omega=\frac{u^\phi}{u^t}=-\frac{g_{t\phi}+ l g_{tt}}{g_{\phi\phi}+l g_{t\phi}} .
$$
the $l$ is the specific angular momentum. The radial velocity $V$ in the rotating frame is
$$
V=\frac{v}{(1-\Omega l)^{1/2}}
$$
where,
$$
v=(-\frac{u_ru^r}{u_tu^t})^{1/2}
$$
Since by definition, $V=1$ on the horizon and sound speed is expected to be
less than unity even for the extreme equation of states, the black hole
accretion has to be supersonic (and therefore sub-Keplerian) on the horizon.
Thus, every flow {\it must deviate from a Keplerian disk} close to a black hole.
The equation for the balance of the radial momentum is obtained from,
$$
(u_\mu u_\nu + g_{\mu\nu}) T^{\mu\nu}_{;\nu}=0,
\eqno{(5)}
$$
which is,
$$
\vartheta\frac{d\vartheta}{dr} + \frac{1}{r \Delta} [a^2- r + \frac{A\gamma^2 B}{r^3}]\vartheta^2
+ \frac{A \gamma^2}{r^6}B
+ (\frac{\Delta}{r^2} + \vartheta^2 )\frac{1}{p+\rho} \frac{dp}{dr}=0
\eqno{(6)}
$$
where,
$$
\vartheta=u^r ,
$$
$$
\gamma^2= [1-\frac{A^2}{\Delta r^4} (\Omega - \omega)]^{-1} ,
$$
$$
B=(\Omega a - 1)^2 - \Omega^2 r^3 .
$$
Here and hereafter we use a comma to denote an ordinary derivative and a
semi-colon to denote a co-variant derivative.
The baryon number conservation equation (continuity equation) is obtained from 
$$
(\rho_0 u^\mu)_{; \mu}=0,
$$
which is,
$$
\noindent {\dot M}=  2\pi r \vartheta \Sigma= 2\pi A \vartheta \rho_0 
\eqno{(7)}
$$
where, $A$ is the cross section area of the flow.
The equation of the conservation of angular momentum is obtained from 
$$
(\delta^\mu_\phi T^{\mu \nu})_{;\nu}=0
$$
or,
$$
\rho_0 u^\mu (h u_\phi)_{,\mu}= (\eta \sigma_\phi^\gamma)_{;\gamma}
\eqno{(8)}
$$
where,
$$
\eta=\nu\rho_0
$$
is the coefficient of dynamical viscosity and $\nu$ is the 
coefficient of kinematic viscosity. When rotation is dominant 
the relevant shear tensor component $\sigma_\phi^r$ is given by (Anderson \& Lemos, 1988;
Straumann, 1991),
$$
\sigma_\phi^r= - \frac{A^{3/2} \gamma^3 \Omega_{,r} \Delta^{1/2}}{2r^5}
\eqno{(9)}
$$
so that the angular momentum equation takes the form,
$$
{\cal L}-{\cal L}_{+}=  
-\frac{1}{\vartheta r^5}\frac{d\Omega}{dr} \nu A^{3/2} \gamma^3 \Delta^{1/2} .
\eqno{(10)}
$$
Where, 
$$
{\cal L}=-h u_\phi
$$ 
and $h$ is the specific enthalpy: $h=(p+\rho)/\rho_0$. 
Here we have modified earlier works of NT73 and Peitz (1994) in that the fluid
angular momentum ${\cal L}$, rather than the particle angular
momentum $-u_\phi$ is used. That way, for an inviscid flow ($\eta=0$)
one recovers ${\cal L}$=constant as in a fluid picture. Similarly, the radial velocity term is
included (eq. 6) and angular momentum is allowed to be non-Keplerian (Eq. 10).
${\cal L}_+$ is the angular momentum on the horizon since shear (as defined 
by Eq. 9) vanishes there. In presence of significant radial velocity, the shear in eq. (9)
is to be replaced by it's full expression, 
$$
\sigma^{\mu\nu}= (u^\mu_{;\beta} P^{\beta\nu} + u^{\nu}_{;\beta}P^{\beta\mu})/2 
- \Theta P^{\mu\nu}/3
$$
where $P^{\mu\nu}= g^{\mu\nu}+u^\mu u^\nu$ is the projection tensor and $\Theta=
u^\mu_{;\mu}$ is the expansion (e.g., Shapiro \& Teukolsky, 1983).
In that case, ${\cal L}_+$ will no longer
be the specific angular momentum of the flow at the horizon, but at
the place where the residual shear effect vanishes. If, on the other hand,  viscosity
is allowed to vanish sufficient rapidly (as we assume in what follows)
on the horizon, then it could be the angular momentum on the horizon as well.

Entropy generation equation is obtained from the first law of thermodynamics
along with baryon conservation equation: 
$$
(S^\mu)_{;\mu}= Q^+-Q^-=[2\eta \sigma_{\mu\nu}\sigma^{\mu\nu}]/T - Q^- ,
\eqno{(11)}
$$ 
which in expanded form, becomes,
$$
\vartheta \Sigma (\frac{dh}{dr} - \frac{1}{\rho_0}\frac{dp}{dr}) =
2\nu \Sigma \sigma_{\mu\nu}\sigma^{\mu\nu} - Q^ -
\eqno{(12)}
$$
where $Q^+$ and $Q^-$ are the heat generation rate (vertically integrated)
and the heat loss rate respectively. $h$ is the specific enthalpy:
$h=(p+\rho)/\rho_0$. Here, we ignore the terms due to radiative transfer.
Using rotational shear as given in eq. (9), the entropy equation takes the form,
$$
\vartheta \Sigma (\frac{dh}{dr} - \frac{1}{\rho_0}\frac{dp}{dr}) =
\frac{\nu \Sigma A^2 \gamma^4 (\Omega_{,r})^2}{r^6} - Q^-
\eqno{(13)}
$$

Along with the equations given above, one needs to solve the relativistic shock
conditions if the flow has a standing shock.
A shock surface is a timelike three surface formally defined as,
$$
\zeta(x^\mu)=0
\eqno{(14)}
$$
across which some of the thermodynamic variables suffer a first order
discontinuity (Taub 1978; Anile \& Russo 1986). The mass and 
the energy-momentum conservation across this discontinuity demands that 
$$
\frac{\partial}{\partial x^\mu} \zeta (\rho u^\mu)_- =
\frac{\partial}{\partial x^\mu} \zeta (\rho u^\mu)_+
\eqno{(15a)}
$$
and
$$
\frac{\partial} {\partial x^\mu} \zeta (T^{\mu\nu})_- =
\frac{\partial}{\partial x^\mu}\zeta (T^{\mu\nu})_+
\eqno{(15b)}
$$
where, the signs $-$ and $+$ denote the quantities before 
and after the shock respectively. The above equations could be written as,
$$
\rho_- u_-^\mu n_\mu = \rho_+ u_+^\mu n_\mu 
\eqno{(16a)}
$$
and
$$
p_-n^\nu + (p_-+\rho_-) (u_-^\mu n_\mu)u_-^\nu  = 
p_+n^\nu + (p_++\rho_+) (u_+^\mu n_\mu) u_+^\nu 
\eqno{(16b)}
$$
Here, $n_\mu$ is the four normal vector component across the shock.

In the next two sections we shall provide explicit examples  
of global solutions of the above set of equations, with or without
shocks. First, we present an approximate method to 
obtain global shock surfaces.

\section{SURFACES OF STRONG SHOCKS IN ACCRETION AND WINDS}

From Eq. (16b) one notices that at a shock surface,
the sum of ram pressure (due to radial motion)
and thermal pressure is continuous across the surface.
Presently, we wish to obtain the nature of this surface assuming that in the
in the strong shock limit one can ignore the
pre-shock thermal pressure compared to the ram pressure.
In this limit, (Novikov \& Thorne, 1973),
$$
\rho_+\sim \frac {\gamma+1}{\gamma-1} \rho_-
\eqno{(17a)}
$$
and 
$$
V_+\sim \frac {\gamma-1}{\gamma+1} V_-
\eqno{(17b)}
$$
where, $\gamma$ is the adiabatic index. Strictly speaking above assumption is 
valid for isothermal flows $\gamma= 1$ only. However, to get a rough approximation
of the global behavior, we assume that this is valid for $\gamma=4/3$ also.
For the post-shock quantities, we ignore radial motion to compute thermal 
pressure, although we include the ram pressure obtained
using Eq. (17b) along with the thermal pressure to obtain the shock surfaces.
(We do not include turbulent pressure here since it involves
an extra parameter, although we believe it's effect is important
as concluded by MLC94.)
Thus, assuming that the post-shock region is predominantly rotating,
one obtains from Eq. (2) (e.g.,Bardeen, 1973), 
$$
\frac{\nabla_i p}{p+\rho} +\frac{\nabla_i u_t}{u_t} 
=\frac{\Omega \nabla_i l}{(1-\Omega l)} .
\eqno{(18)}
$$
In this limit the equation could be integrated for any power law
angular momentum distribution (C85),
$$
l=c_\phi \lambda^{n_\phi}
\eqno{(19)}
$$
provided a barotropic equation of state $p=p(\rho)$ is chosen. Here,
$c_\phi$ and $n_\phi$ denote two positive constants with $0.0\lsim
n_\phi \lsim 0.5$ in the disk region. This latter constant actually
represents the viscosity of the flow: for a small viscosity,
angular momentum is nearly constant and $n_\phi \sim 0$. For high viscosity,
the angular momentum is nearly Keplerian, $n_\phi \sim 0.5$.
Also, the von-Zeipel parameter (e.g., C85; Chakrabarti, 1991) $\lambda$ is given by,
$$
\lambda^2=\frac{l}{\Omega} = -\frac{l g_{\phi\phi} + l^2 g_{t \phi}}
{g_{t \phi} + l g_{tt}}
\eqno{(20)}
$$
We choose a flow with equation of state $p=K\rho_0^\gamma$,
where $K$ is a constant.  Using a distribution given by Eq. (19) one obtains the
enthalpy $h=(p+\rho)/\rho_0=1/(1-na_s^2)$ (with sound speed
$a_s$ defined by, $a_s^2=\frac{\partial p}{\partial\rho}|_s$ at constant
entropy $s$. $n=[\gamma-1]^{-1}$ is the polytropic index) 
distribution from Eq. (18) as,
$$
h u_t (1-\Omega l)^\alpha = {\cal E} ,
\eqno{(21)}
$$
where,
$$
\alpha=\frac{n_\phi}{2n_\phi-2} .
\eqno{(22)}
$$
Equation (21) represents the general relativistic Bernoulli
equation. Here, due to the variation of angular momentum, the specific
energy $h u_t={\cal E}_0$ is not conserved.
Only in an inviscid flow, i.e., for $n_\phi=0$, ${\cal E}={\cal E}_0$.
The time component of velocity $u_t$, which is also
the specific binding energy, could be obtained from Eq. (3) as,
$$
u_t=\left [\frac{g_{t\phi}^2 - g_{tt} g_{\phi\phi}}
{(g_{\phi\phi}+ l g_{t\phi})(1-\Omega l)(1-V^2)} \right]^{1/2}
\eqno{(23)}
$$

Before the incoming matter hits this predominantly rotating flow
configuration, its thermal pressure was negligible, and assuming the 
same angular momentum distribution (Eq. 19) in the pre-shock flow
one obtains the ram pressure distribution from the following equation,
$$
u_t (1-\Omega l)^\alpha = {\cal E} ,
\eqno{(24)}
$$
Here we have chosen specific enthalpy $h=1$ for a cold inflow in the pre-shock
region. One need not 
have an angular momentum distribution as given by Eq. (19) in the pre-shock
region, since the barotropic flow is no longer pre-dominantly rotating, and therefore
the von-Zeipel theorem (Tassoul, 1978) need not be strictly imposed. But
we choose it anyway so as to obtain a purely compressible shock wave
where the angular momentum remains continuous across the shock. 
In any case, our conclusions are not found to be qualitatively
different if the factor  $(1-\Omega l)^\alpha$ is dropped instead.

At the shock surface, the pressure balance equation (16b) takes the form:
$$
p_+ + (p_+ + \rho_+) V_+^2/(1-V_+^2) \sim (p_-+\rho_-) \frac{V_-^2}{1-V_-^2} 
\eqno{(25)}
$$
Note that we have not taken the projection of velocity along
the normal to the shock surface here. This is because in the strong
shock limit that we are considering here, the surfaces of constant
ram pressure outside the shock and the surfaces of constant
thermal pressure inside the shock would have similar functional dependence
through $u_t (r,\ \theta)$. Thus
locally the ram pressure gradient is parallel to the thermal pressure
gradient at each point on the shock surface.

Combining Eqs. (17a,b) and Eq. (25) we obtain the pressure balance condition 
at the shock as,
$$
\frac{n a_s^2}{1-na_s^2} \sim  
\frac{2\gamma}{(\gamma +1)^2} \frac{V_-^2}{1-V_-^2}.
\eqno{(26)}
$$
Figs. 1a-b show examples of the shock surfaces 
(S, solid curves) and the contours (short dashed curves) of constant thermal 
pressure inside the shock surface while
the contours of constant ram pressure outside the shock surface.
Contours are in axial distance $R=rsin \theta $ vs. vertical distance 
$Z=r cos \theta$  (namely, in meridional) plane.
The parameters we choose are $a=0.95$, $c_\phi=2.3$, $n_\phi=0.05$
and $\gamma=4/3$. We also show the funnel surface defined as the
zero pressure surface (F) and the centrifugal barrier (CF)  defined
by the surface of maximum pressure ($\nabla p=0$). In a predominantly rotating
flow (Eq. 18), the vanishing pressure gradient implies that only 
competing forces at CF are the centrifugal, gravitational and Coriolis' forces
(in Newtonian terminology). Thus, the centrifugal barrier
resembles the Keplerian distribution, so to speak.
In Fig. 1a, we choose ${\cal E}=1.001$ constant at the outer boundary
at $R=50$. In this case, the shock surface that begins on the equatorial
plane bends inside. This is because the radial velocity becomes smaller
as the matter approaches the centrifugal barrier and the flow has to move in farther
to match the thermal pressure. The segment of the shock surface
inside the centrifugal barrier is not physical for accretion flow,
as the flow has to move outwards to form them. This segment is
however, physical for winds blowing outwards. Thus both the winds and 
accretion can have shock waves simultaneously in different
regions of the flow. The
segment outside the centrifugal barrier is equivalent to the shock at 
$x_{s3}$ in the notation of C89 and the segment inside the centrifugal
barrier is equivalent to $x_{s2}$ in the notation of C89. As shown
in Chakrabarti \& Molteni (1993), in a polytropic flow, $x_{s3}$ is stable for
accretion while $x_{s2}$ is stable for winds.

In Fig. 1b, we artificially choose an outer boundary condition where the
specific energy of the flow increases as one goes away from the equatorial
plane. We choose: ${\cal E} (\theta)=0.03 cos \ \theta + {\cal E}(\theta=
\pi/2)$, where $\theta=0$ coincides with the vertical axis.
Motivation for this choice is that in an astrophysical situation
(as well as in a typical numerical simulation)
infall radial and sound velocities are constants at the outer
boundary. At a higher elevation, the potential energy decreases, so the
net specific  energy increases. The flow forms a shock farther away from the axis
as less centrifugal force is required to balance the weakened gravity.
In all the numerical simulation results reported (HSW85, HSW85, H84, MLC94, 
MRC96) such shock waves which bend outwards are seen. Indeed in H84, several examples
of shocks which bend back towards the vertical axis at a even higher
latitude are reported exactly as shown in Fig. 1b. After the flow
passes through this oblique shock, the energy may be randomized, 
and another shock of the type shown in Fig. 1a may be formed. These
are seen in MLC94 and MRC96 as well.  

Note that in the strong shock limit that we discuss here,
the outer sonic surface lies at infinity, while the inner sonic 
surface lies arbitrarily close to the horizon. This is simply because the
flow is chosen to be very cold $a_s=0$ in the pre-shock flow
and almost static $V\sim0$ in the post-shock flow. When these constraints are
relaxed, as in our analysis in the next section, the outer sonic point
{\it moves in} from infinity to a few tens to a few hundreds of
Schwarzschild radii,
while the inner sonic point {\it moves out} to roughly
one to three Schwarzschild radii as see in pseudo-Newtonian geometry (C89, C90).

\section{SHOCKS AND SHOCK-FREE SOLUTIONS IN A CONICAL WEDGE FLOW}

In the last section we concentrated on strong shock solutions.
Presently, we study solutions of the transonic 
flows around a rotating black hole in full generality. 
General procedures for obtaining such solutions using pseudo-Newtonian potential
have already been in numerous works, and we shall not repeat them here
(see, C89, C90, C96b). We use this weak viscosity
limit of the equations of Section 2
to obtain the zeroth order distribution of velocity and sound speed
etc. Subsequently, we compute the effects of viscosity on this zeroth order distribution.
Computation of heating and cooling effects are also done in the same way.

Before proceeding further, we would like to argue that the conical flows have 
all the properties similar to other types of thin flows, such as flows in vertical 
equilibrium and flows of constant thickness. Different models
can really be mapped onto each other by varying
the polytropic constant appropriately. To show this, we 
consider a gram of matter of specific energy ${\cal E}$ and
specific angular momentum $l$ in Newtonian geometry:
$$
{\cal E}=\frac{1}{2}V^2 + n_i a_s^2 + \frac{l^2}{2r^2} - \frac{1}{r} .
\eqno{(27a)}
$$
Using the equation of state
$p=K\rho^\gamma$ (where $\gamma=1+1/n$ is the adiabatic index, and
$K$ is a  constant), the mass accretion rate on the equatorial plane
$$
{\dot M}_i= \rho  V r^q
\eqno{(27b)}
$$
could be re-written as,
$$
{\dot{\cal M}_i}=V a_s^{\nu_i} r^{q_i}
\eqno{(27c)}
$$
where ${\dot{\cal M}_i}$ is an entropy dependent quantity $\propto
{\dot M}_i K^{n_i}$. Here $V$ is the radial velocity and the 
subscript $i$ differentiates one model from another.
For spherical or conical flows, ($i=1$, say), $q_1=2, \ \nu_1=2 n_1$, while
for a thin flow of constant height ($i=2$, say), $q_2=1,\ \nu_2=2n_2$ and
finally, for a thin flow in vertical equilibrium ($i=3$, say),
$q_3=5/2, \ \nu_3=2n_3+1$. It can be easily shown that the properties
at the sonic points remain identical in all these three models, provided,
the polytropic indexes are modified according to the relation,
$$
\frac{2n_1-3}{4}=\frac{2n_2-1}{2}=\frac{2n_3-3}{5}.
\eqno{(28)}
$$
Due to non-linearity, a cleaner relation such as this is impossible in general
relativity. But we expect the situation to be similar. Therefore,
we feel confident that a conical wedge shaped flow can capture the
physical process as any other thin flow model. 

In the case weakly viscous flow, Eq. (6) could be integrated to give the
energy conservation,
\noindent 
$$
hu_t = \frac {p+ \rho} {\rho_0} u_t = {\cal E} .
\eqno{(29a)}
$$
The baryon number conservation relation remains the same as Eq. (7), with
the flow cross-section area to be
$A \propto 2\pi r^2 sin \ \theta \delta \theta$ where $\delta \theta$
is the  angular width of the flow around the angle  $\theta$ with the
vertical axis (in C90, a simpler accretion rate was used to obtain solutions in Kerr geometry)
Thus,
$$
{\dot M} = 2\pi \rho_0 r^2 u^r = {\rm constant}
\eqno{(29b)}
$$
and finally, the angular momentum conservation equation (Eq. 8) becomes simply,
$$
hu_\phi =\frac {p + \rho}{\rho_0} u_\phi = -{\cal L}.
\eqno{(29c)}
$$
We solve these equations by using the standard sonic point analysis.
First rewrite the energy and mass conservation equations as
(e.g., Shapiro \& Teukolsky, 1983),
$$
{\cal E}= \frac{1}{1-na_s^2} \frac{1}{(1-V^2)^{1/2}} F(r,\theta)
\eqno{(30a)}
$$
and
$$
{\dot {\cal M}} = \left ( \frac{a_s^2}{1-na_s^2} \right )^{n}
\frac{V}{(1-V^2)^{1/2}}  G(r,\theta)
\eqno{(30b)}
$$
where,
$$
F(r,\theta)=
\left [ \frac{g_{t\phi}^2-g_{tt}g_{\phi\phi}}{(g_{\phi\phi} +
l g_{t\phi})(1-\Omega l)} \right ]^{1/2}
$$
and
$$
G(r, \theta)= \frac{r^2 sin \theta }{g_{rr}^{1/2}}.
$$
Here, we have neglected $\theta$ component of velocity $u_\theta$ 
compared to the other two spatial components $u_r$ and $u_\phi$.
This assumption need not be valid, especially in the immediate post-shock
region, where $u_r$ itself becomes small enough. However, we believe
that we can capture the basic physics of transonic flows this
approximation. The quantity ${\dot {\cal M}}$ will be called the entropy 
function (earlier, in C89 \& C90, the terminology of `accretion rate' 
was used for this quantity, though it was distinguished from the mass flux ${\dot M}$)
as ${\dot {\cal M}} \propto K^n {\dot M}$.
It usually assumes two different values at two saddle type sonic points and 
the one with the smallest ${\dot {\cal M}}$ 
joins the horizon at infinity (C90). In a flow which contains
shocks, matter first passes through this transonic solution of lower entropy,
and then through a shock and finally through the sonic point
which requires a higher entropy. 

Differentiating equations (30a) and (30b) with respect to $r$ (for a given $\theta=\theta_0$)
and eliminating terms involving derivatives of $a_s$, we obtain,
$$
\frac{dV}{dr} = \frac {N}{D}= \frac{c_1 a_2 - c_2 a_1}{b_1 a_2 - b_2 a_1}
\eqno{(31)}
$$
where,
$$
a_1=\frac{2na_s}{1-na_s^2},
\eqno{(32a)}
$$
$$
a_2=\frac{2n}{a_s(1-na_s^2)},
\eqno{(32b)}
$$
$$
b_1=\frac{V}{1-V^2},
\eqno{(32c)}
$$
$$
b_2=\frac{1}{V(1-V^2)},
\eqno{(32d)}
$$
$$
c_1=\frac{1}{F(r,\theta_0)}\frac{dF(r,\theta_0)}{dr},
\eqno{(32e)}
$$
$$
c_2=\frac{1}{G(r,\theta_0)}\frac{dG(r,\theta_0)}{dr},
\eqno{(32f)}
$$

From the vanishing condition of $D$ and $N$ at the sonic points,
one obtains the so called sonic point conditions as,
$$
V_c=a_{s,c}
\eqno{(33a)}
$$
and
$$
V_c^2=\frac{c_1}{c_2}|_c
\eqno{(33b)}
$$
Here, the subscript `c' refers to the quantities evaluated at the
sonic point $r=r_c$. 

In the study of inviscid accretion flows, one supplies the conserved
quantities, such as $l$, ${\cal E}$, etc. to obtain a complete solution.
For a given angular momentum $l$, the remaining
unknowns are $V (r)$, $a_s (r)$. But one requires
only one extra boundary condition, e.g. ${\cal E}$, since
two sonic point conditions (33a,b) introduce only one extra
unknown $r_c$. Both the quantities $l$ and ${\cal E}$ could be
functions of $\theta$ in general and one has to supply another constant
equivalent of the Carter constant (Chandrasekhar, 1983)
to describe flows in the region away from the equatorial plane. However,
since we let $u_\theta=0$ everywhere, we do not have 
such a choice any more. Thus, the supply of the
specific energy ${\cal E}$ and the specific angular momentum $l$
is sufficient to obtain a complete solution from the horizon to infinity.
Note that by definition $\Omega$ is the same as $\omega$ on the horizon.
At the shock, apart from the continuity of the energy and the mass flux,
one must also satisfy the relativistic momentum balance condition, Eq. 16(a-b).

\subsection{Classification of the Parameter Space}

A solution, with or without a shock, can be completely described
by the choice of  either ($l,{\dot {\cal M}}$) or ($l,{\cal E}$)
pair. As in a flow in a pseudo-Newtonian geometry (Fig. 2 and Fig. 4 of C89), 
one can classify the entire parameter space in ($l,{\dot {\cal M}}$) or ($l,{\cal E}$)
plane in terms of whether or not the flow has one or more sonic points
and shocks. Fig. 2 shows such a classification (for $a=0.95$)
for the flow on the equatorial plane ($\theta=\pi/2$).
The parameter space is sub-divided into nine regions. Flows with parameters from the
region $N$ has no transonic solution. 
The solutions from the region `O' have only the outer sonic point.
The solutions from the regions $NSA$ and $SA$ have two `X' type sonic points
with the entropy density $S_o$ at the outer sonic point {\it less} than the
entropy density $S_i$ at the inner sonic point. However, flows from $SA$
can pass through a standing shock (See, Figs. 3 and 4 below) as the Rankine-Hugoniot
condition is satisfied. The entropy generated at the shock is exactly
$S_i-S_o$, which is advected towards the black hole to enable the flow to pass
through the inner sonic point. Rankine-Hugoniot condition is not satisfied
for flows from the region $NSA$. Numerical simulation indicates
(Ryu, Chakrabarti \& Molteni, 1997) that the solution is very unstable
and show periodic changes in shock locations.
The flow from the region $SW$ and $NSW$ are very similar to those from
$SA$ and $NSA$. However, $S_o \geq S_i$ in these cases.
Shocks can form only in winds from the region $SW$. The shock condition is not
satisfied in winds from the region $NSW$. This may make the $NSW$ flows
unstable as well. However, stable accretion can take place through the inner sonic
point. A flow from region $I$ only has the inner sonic
point and thus can form shocks (which require
the presence of two saddle type sonic points) only if the inflow
is already supersonic due to some other physical processes. See C90
and Chakrabarti (1996c) for all the topologies of the flow.

The flows from regions $I^*$ and $O^*$ are interesting in the sense
that each of them has two sonic points (one `X' type and one `O' type)
only and neither produces a globally complete solution (in the
inviscid case, at least). The region $I^*$
has an inner sonic point but the solution does not extend subsonically
to a large distance. The region $O^*$ has an outer sonic point, but the
solution does not extend supersonically
to the horizon! In both the cases a weakly viscous
flow is expected to be unstable. When a significant viscosity is added, the closed
topology of $I^*$ is expected to open up (C90, C96b) and then the 
flow can join with a Keplerian disk.
Clearly, advective disks from this region of parameter space
will have no standing shocks independent of what polytropic index is used
unless some extra energizing mechanism (such as flares) in the sub-Keplerian
region raises the specific energy of the flow to bring it to the
parameter space marked `SA'.

\subsection{Behavior of Shock Locations on the Equatorial Plane}

First, we study the properties of the shock waves on the equatorial
plane ($\theta=\pi/2$). Our goal is to see if the shocks
in a true black hole geometry display similar 
properties as those in pseudo-Newtonian geometry.
With a pseudo-Newtonian geometry, it was noted that
for given flow parameters, there are two locations $x_{s2}$
and $x_{s3}$ where shocks form in between two saddle type sonic points. 
Later, it was found (Chakrabarti \& Molteni, 1993; also see Nakayama, 1992
for isothermal flows) that $x_{s3}$ is
stable in accretion and $x_{s2}$ is stable in winds.
In Fig. 2, we show shock locations in four cases,
(a) $a=0,\  l=3.5$, (b) $a=0.5,\  l=3.0$, (c) $a=0.95,\  l=2.3$
and (d) $a=0.95,\ l=2.1$ as a function of the specific energy of the flow. 
Each `Persian bottle' shaped set is marked with the above
parameters. Each set contains four curves, two dashed and two solid. 
The dashed curve on the lower left ($a_2$)
corresponds to the unstable shock $x_{s2}$ of accretion flow
and the solid curve on the upper left ($a_3$) corresponds to the stable
shock $x_{s3}$ of accretion flow. The solid curve on the lower right ($w_2$)
corresponds to the stable shock $x_{s2}$ of the wind flow
and the dashed curve on the upper right ($w_3$) corresponds to the unstable
shock $x_{s3}$ of wind flow.  This description is applicable to
{\it each} set. First notice that the shocks
are possible on the equatorial plane for a certain range of
energy values only. The general behavior of the shock
location is {\it exactly} same as in pseudo-Newtonian geometry
(C89; C90; Chakrabarti \& Molteni, 1993). What is more, the range of energy
for which shocks are possible in a rapidly rotating black hole
is much higher compared to that around a non-rotating black hole.
On the average, the shock locations are 
closer to the horizon for flows around a rotating black hole.

\subsection{Shear, Heating and Cooling in a Viscous flow}

Since the nature of the transonic solution including the formation of shocks
depends on angular momentum distribution, which in turn depend
very strongly upon viscosity cooling and heating, it is instructive to
study their behavior. Presently, we study then in the weak viscosity
limit, where the velocity and density distributions are chosen from the
inviscid flows while computing shear, heating and cooling.  The complete
solutions of the viscous transonic flow which self-consistently
compute these quantities along with the flow variables (such as, velocity,
density,  etc.) will be dealt with elsewhere.

Around a Newtonian star, shear component due to purely
rotational motion is defined as $\sigma_{r\phi}= r d\Omega/dr$ which is
always positive (e.g, $3/2 \Omega$ for a Keplerian disk), and
hence viscous stress is always negative. That is,
angular momentum is always transported outwards. However, Anderson \& Lemos (1988)
pointed out that the shear can change sign near the horizon and they
demonstrated this using velocity profiles of a cold radial
flow below marginally stable orbit. Below, we show that at least for a
weakly viscous flow shear indeed changes its sign close to a black hole.

In Fig. 4, we show the behavior of shear components $\sigma^r_{\phi}$
for a prograde and a retrograde flow. In the upper left panel, we show the Mach
number variation of the flow with $a=0.95, \ l=2.3,\ {\cal E}=1.001$.
The arrowed path shows that two transonic solutions are `glued' together
by the vertical shock transition (at $a_3$ or $x_{s3}$ in our notation)
where Rankine Hugoniot condition is
satisfied. On the upper right panel, We present 
$\sigma^r_{\phi p}$ and $\sigma^r_{\phi}|_{rot}$ which are the
general shear and the rotational shear (Eq. 9) for the supersonic branch passing 
through the outer sonic point (subscripts `p' and `b' are for supersonic
and subsonic branches, respectively). For comparison, we also show
$\sigma^r_{\phi b}$ (general shear for the subsonic branch passing 
through the outer sonic point) and $d\Omega/dr$ (which is independent of
any branch). One arrives at several important conclusions: First,
the effect of radial motion is significant in changing the shear
stress. The rotational shear $\sigma^r_{\phi{|_{rot}}}$ as well as 
$\sigma^r_{\phi b}$ where the radial velocity is negligible, vanish at 
the horizon, while the $\sigma^r_{\phi p}$ does not. Indeed,
$\sigma^r_{\phi p}$ is a very large negative quantity, and 
therefore in presence of even a weak viscosity it can transport angular momentum
inwards. This does not automatically imply that the angular momentum distribution 
will develop a minimum (such as in the Keplerian distribution), see Fig. 5 below.
In reality, this inward transport may try to nullify a steeper 
distribution with a positive slope and makes it as flat as possible
near the horizon. The lower left panel of Fig. 4 gives the variation of Mach 
numbers for a retrograde flow with $a=0.95, \ l=-4.0, \ {\cal E}=1.0045$. 
Note that both the subsonic and supersonic branches behave qualitatively
similarly, although that from the supersonic branch ($\sigma^r_{\phi p}$)
becomes extremely large positive quantity before turning back
to become a large negative just outside the horizon. 
We have not plotted the shear distribution for the post-shock
branches passing through the inner sonic points, as their
behaviors are similar. What is very much clear is that a simple
Shakura-Sunyaev (1973) type viscous stress $\propto -\alpha p$
with a constant $\alpha$ will not be very accurate, specially close to the
black hole, since the pressure $p$ is always positive  while actual
stress changes sign!

The reversal of the shear close to a black hole is no mystery. 
In general relativity, all the energies couple one another. 
It is well known that the `pit in the potential' of a black hole 
is due to coupling between the rotational energy and
and the gravitational energy (see, Chakrabarti, 1993 and references therein).
As matter approaches the black hole, the rotational
energy, and therefore `mass' due to the energy increases which is also
attracted by the black hole. This makes gravity
much stronger than that of a Newtonian star. When the black hole
rotates, there are more coupling terms (such as that arises out of
spin of the black hole and the orbital angular momentum of the matter)
which either favor gravity or go against it depending on whether
the flow is retrograde or prograde  respectively.
This is the basic reason why retrograde and prograde flows
display different reversal properties.

Given that the shear is a very large negative number close to a horizon,
for a physical flow, one must have the viscosity vanishing on the
horizon in order to have finite angular momentum or the viscous dissipation.
Fig. 5 shows the distribution of angular momentum (upper panel)
and the heating and cooling rates (lower panel) in the pre-shock 
(with subscript `pre') and post-shock (with subscript `post') solutions
for the prograde flow presented in Fig. 4 above. We also present
the Keplerian distribution ($l_K$) in the upper panel for comparison. 
We assume ion viscosity (Spitzer, 1962) with a  factor $(1-r_+/r)^\beta$ in 
order that it vanishes on the horizon sufficiently rapidly (we use $\beta=1$ here). 
The same variation is used to compute viscous heating rate: 
$$
Q^+=2\eta\sigma_{\mu\nu}\sigma^{\mu\nu}=-\frac{4\eta}{\Delta}\sigma_{r\phi}\sigma^r_\phi
(\Omega^2 g_{\phi\phi} +2\Omega g_{t\phi}+g_{tt})
\eqno{(34)}
$$
Note that the quantity in the parenthesis has to be negative in order to obtain
positive heating. This is indeed always the case for any realistic flow
(e.g., Shapiro \& Teukolsky, 1983). For the cooling, we simply use bremsstrahlung
process. In presence of a Keplerian disk farther out, 
Comptonization  of the intercepted photons
is likely to dominate (Chakrabarti \& Titarchuk, 1995). We 
chose $M_{bh}=1M_\odot$ and ${\dot M}={\dot M}_{Eddington}$ for
illustration purposes (as we are interested in general behavior, and not the absolute
magnitude). In the upper panel, dotted curve is the angular momentum distribution in the 
preshock branch and solid curve is that in the post-shock branch. The dotted vertical
arrow is at the shock location. In the lower panel the solid curves are viscous 
dissipation rate and the dotted curves are the bremsstrahlung cooling rate. Clearly, since 
the post-shock flow is hotter, ion viscosity is higher. This causes a rapid transfer
of angular momentum as well as a higher heating and cooling rates in the post-shock region. 
In reality, when flow velocity, sound speed etc. are
computed simultaneously, the effect becomes much weaker,
and the piling up of angular momentum would either move the shock
outward, or the shock may disappear completely as has been observed in 
recent numerical simulations (see, C96a and References therein).
Most of the heat generated in the present case is carried away towards the
black hole, since bremsstrahlung is not efficient enough to cool the flow. 
This is similar to the inviscid case where energy
remained conserved (C89), or in viscous pseudo-Newtonian flows 
with inefficient cooling (C96a, C96b).

Instead of using ion viscosity, if one used parametrized
kinematic viscosity such as $\nu \sim \alpha c_s r$ where, $c_s$
is the isothermal sound speed and $\alpha$ is the Shakura-Sunyave (1973)
viscosity parameter, then also we would have exactly the
same behavior as above.  However, for finite heating and
angular momentum transport rate, $\alpha$ must vanish
`sufficiently rapidly' on the horizon. This suggests that
it is essential to have a fundamental change in viscous
processes very close to the horizon. A constant $\alpha$
description may be adequate for `predominately rotating flows',
however, since the rotational shear $\sigma_\phi^r|_{rot}$
vanishes on the horizon (Fig. 4). Such predominantly
rotating flows do not satisfy the boundary condition $V=1$ on the
horizon  and we prefer to conjecture that the turbulence and viscosity
be completely damped out on the horizon instead (`tranquil inflow conjecture').

\subsection{Shock Location away from the Equatorial plane}

If one considers conical flows away from the equatorial plane,
the shocks are generally expected to recede from the axis. This is
simply because away from the plane the gravitational force 
diminishes so that less centrifugal force is required to form a
shock. For illustration, we consider the same three cases as in the above
sub-section, but we choose a particular energy ${\cal E}=1.004$
for all the three cases. From Fig. 3, we note that shock
solutions on the equatorial plane exist for all the three cases at this energy. 
In Fig. 6, we show the variation of shock location (dotted curves 
marked with S) with height in the $R-Z$ plane. For comparison, we show the
funnel wall (solid curves marked with F) and the centrifugal barrier 
(long dashed curves marked with CF).
The outer most curve in each set is for $a=0$, the middle curve
is for $a=0.5$ and the innermost curve is for $a=0.95$. The location
of the shock is more distant from the vertical axis at a higher elevation.
The shock strengths (measured by the ratio of pre-shock to
post-shock Mach numbers $M_-/M_+$) fall sharply
with increasing latitude. The shocks in accretion flows typically exist
in the range $\theta_{max} \geq \theta \geq \theta_{min, a}$ 
(see Fig. 7 below), where $\theta_{max}$ is not necessarily $\pi/2$. In other 
words, even for parameters which do not allow shocks on the equatorial
plane, the flow may have shocks starting at an angle $\theta_{max}<\pi/2$ away from
the  equatorial plane and extending up to $\theta_{min,a}$. This happens when
the energy is less than the lower limit of energy required
for the shock in the equatorial plane
(which is different for different cases as depicted in Fig. 2 and Fig. 3). For instance,
flows from the region $NSA$ (Fig. 2) would have shocks at a higher elevation, though
on the equatorial plane shocks were not allowed.
Beyond $\theta_{min,a}$, the flow can have stable shock
$x_{s2}$ only in winds up to some other $\theta_{min,w}$. These shocks
form very close to the centrifugal barrier near the black hole,
and for clarity we do not draw them.
These shocks also bend away from the vertical axis at higher elevations.
Similar to accretion flows, winds can develop shocks at higher
elevation when the energy is more than the highest energy required for shocks on the
equatorial plane (which is different in different cases as depicted in Fig. 2 and Fig. 3).
Thus flows from region $NSW$ (Fig. 2) would have shocks in winds at higher elevation.
Again for clarity we have not plotted outer and inner sonic surfaces.
These `annular shocks' are observed in simulations of Ryu et al. (1995) and Ryu,
Chakrabarti and Molteni, 1996). The outer sonic surface is very close to spherical
(as it is hundreds of Schwarzschild radii away, and therefore insensitive 
to rotational effects) moving in slightly with elevation. The inner sonic surface located 
close to the horizon moves inwards with elevation at a comparatively faster rate.

In Fig. 7 we compare the entropy function ${\dot{\cal M}}$ 
for all the three cases discussed in \S 4.2
as a function of $\theta$ ($\theta=90$ corresponds to the equatorial plane). 
For a given mass accretion rate ${\dot M}$ for all the three cases,
these curves measure the entropy of the flow and the amount of entropy
that is generated at the shock.
The three dashed curves, roughly overlapping each other
are the entropy functions at the outer sonic surfaces. These
sonic surfaces being very far away, their properties are almost independent
of the state of the black hole or the (sub-Keplerian) 
angular momentum of matter. These are the entropies of the pre-shock
flow since no entropy is generated before the flow hits the shock. 
The solid curves represent the
entropy of the post-shock flow which pass through the inner sonic surfaces. Clearly,
the flow around the rapidly rotating black hole is the hottest
and has the highest entropy. The difference in entropy between the
outer and the inner sonic surfaces must be generated at the 
shock. Therefore, a higher jump (at a given angle) in entropy represents
a stronger shock. Note that at some $\theta_{min,a}$ the solid and dashed
curves intersect. That is the location (angle) where the weakest shock 
(with strength unity) forms. For $\theta_{min,a} <\theta \leq \pi/2$, the shock in accretion
flow can form, since it has to {\it generate} entropy at the shock.
Beyond that the entropy at the outer sonic surface is lower than that from the
inner sonic surface, and the shocks are formed only in winds. Accretion flow
with these parameters will not have shocks. This process continues till 
$\theta_{min,w}\sim 60$ degrees. Beyond that angle, shocks will not form in winds as well.

\section{CONCLUDING REMARKS}

In this paper, we studied the general behavior of the 
transonic solutions, with or without shocks,
in  two dimensional accretion and winds. We employed fully
general relativistic equations for viscous transonic flows for this purpose. 
We show that the behavior of shock waves on the equatorial
plane is very similar to what had been found 
using pseudo-Newtonian potential. However, for rotating 
black holes, the range of parameter space which allows the
formation of shock waves is much larger. Similarly,
flows away from the plane form shocks more easily.
These studies therefore imply that shocks in  quasi-spherical accretion flows
are perhaps more generic. Furthermore, it can be easily shown,
by employing a perturbation of pressures on both sides of the shocks,
that these shocks are stable as well (Chakrabarti \& Molteni, 1993).
Both of these deductions from purely analytical considerations
as presented in this paper are borne out by every numerical simulations
to date (HSW84, HSW85, H84, MLC94, MRC96). These simulations always 
find very large scale stable oblique shocks in a black hole
accretion. In case when the flow is bound, namely ${\cal E} <1$, 
the flow does not have outer sonic point, and  does not extend to a
large distance (region $I^*$ in Fig. 2). Only when viscosity is added, the 
closed topology around `O' type sonic point 
opens up to join a Keplerian disk farther out. Shocks can form on the
equatorial plane in this case only when the inflow is already supersonic,
though away from the plane the flow may have ${\cal E}\geq 1$ and 
stable shocks will form.
This considerations were used to explain the spectral
properties of black hole candidates (Chakrabarti \& Titarchuk, 1995).
In the advective corona ${\cal E} \geq 1$ condition could also be achieved by
energy deposition in the corona through magnetic flares (for instance)
a method which is widely accepted in high energy astrophysics.

Our `brute force'  method of global shock study (Section 3) solves the 
mystery that was first noticed by Fukue (1987) and 
subsequently by C89 as to the multiplicity in formal shock locations. 
Since the transonic flow we consider are necessarily
non-Keplerian, they typically tend to have a pressure maximum 
exactly as in a thick accretion disk. As a result,
the shock condition is satisfied once on each side of the maximum. Our method
also reveals why the outer shock is stable in accretion and inner shock is
stable in winds. In both the cases we see that the thermal pressure
must increase downstream. In fact, for a highly viscous flow the pressure
maxima and consequently the shocks are absent (C96b).

Since general topology of the solutions depend on angular momentum distribution,
we also studied the behavior of shear for a transonic flow. We conclude that
shear is not monotonic close to a black hole. In fact it may
reverse several times due to coupling of various 
energies in general relativity. Because of this we believe that neither
$\alpha$ viscosity prescription, nor computation based on predominantly
rotating flow (rotational shear, Eq. [9]) should be adequate. We find that
heating is very strong close to the black hole, and the generated heat 
must be advected away with the flow if the cooling is not very
efficient. This is 
similar to inviscid case (e.g. C89), where the entire energy and entropy at the
shock was advected by the optically thin flow. The post-shock solution 
is observed to transport angular momentum more rapidly and as a result
the shock may either move away from the black hole, or it may disappear
if the viscosity is high enough. These results agree with our earlier
computations in pseudo-Newtonian flows. 
 
Since black holes have no hard surface one would imagine that
matter would enter through the horizon without decelerating. But it seems
that the centrifugal barrier by even a small amount of angular momentum
in the flow causes the matter to pile up behind the barrier
over a large scale, not just near the equatorial plane. Indeed,
we found that the probability  of formation of the shocks
away from the plane is much higher, simply because the 
barrier is definitely present away from the plane (due to weaker gravity)
even when it is absent on the equatorial plane (such as when the
angular momentum of the flow is everywhere sub-Keplerian close to the
black hole). Our present findings support the view that the black hole accretion 
models may need to include shock waves (at least for flows at higher elevation) as they
provide a complete explanation of the observed steady state as well as time
dependent behaviors (e.g., Chakrabarti \& Titarchuk, 1995;
Ryu, Chakrabarti \& Molteni, 1997; Crary et al, 1996).
For instance, according to our model, quasi-periodic oscillations of black hole candidates
should have frequencies of the order  of $1/(4x_s^{3/2})$ (Molteni,
Sponholz \& Chakrabarti, 1996) and therefore can vary from a few millihertz
to kilohertz range depending on shock locations (Fig. 3) with modulation
amplitude as much as $100$ percent. This roughly agrees with the observed
frequencies and amplitudes.

The author is thankful to his colleagues D. Ryu and D. Molteni
for excellent numerical simulations which partly motivated 
him to complete the present work. He is also thankful to J. Hawley
for giving him a copy of his thesis ten years ago, which 
contained insightful illustrations which were helpful in 
understanding two dimensional transonic flows. He thanks
J. Peitz, R. Khanna and Kip Thorne for discussions.

\clearpage

\begin{center}
{\bf FIGURE CAPTIONS}
\end{center}
\begin{description}

\item[Fig.~1a-b] Funnel wall (F), centrifugal barrier (CF) and
the strong shock surface  (S) in a flow which is cold in the pre-shock
region and predominantly rotating in the post-shock region. The 
dashed contours inside the shock are those of thermal pressure plus
ram pressure while those outside are those of ram pressure. 
In regions outside the centrifugal barrier, the shock is stable
in accretion while that inside is stable in winds.
In (a) energy is constant at a constant von-Zeipel cylindrical surface, while
in (b) energy is increasing with latitude to mimic boundary
conditions of numerical simulations.

\item[Fig. ~2] Classification of the entire parameter space spanned by 
the specific energy and angular momentum in terms of the number of sonic
points and the presence or absence of shocks. See text for details.

\item[Fig.~3] Stable (solid) and unstable (dashed) shock locations
in accretion ($a_2$, $a_3$) and winds ($w_2$ and $w_3$)
as a function of the specific energy energy of the flow for four
black holes: $a=0, \ l=3.5$; $a=0.5, \ l=3.0$; $ a=0.95, \ l=2.3$; $a=0.99, \ l=2.1$.

\item[Fig.~4] Variation of Mach numbers (left panels) and shear components (right
panels) in a prograde (upper panels) and retrograde (lower panels) flow. The shock
transition is shown as a vertical arrow. Shear components are not monotonic as in a 
Newtonian flow, but reverses near the black hole. $\sigma^r_\phi |_{rot}$ is for
rotational shear (assuming the flow is predominantly rotating), 
whereas $\sigma^r_{\phi p,\phi b}$ are for general shear which includes the radial motion. 

\item[Fig.~5] Variation of the angular momentum distribution 
(upper panel) and the heating and cooling rates (lower panel) 
in presence of ion-viscosity and bremsstrahlung effects respectively. Most of the
heat generated by the viscous transonic flow in this case
is carried by the flow inwards as the bremsstrahlung is inefficient in radiating it away.

\item[Fig.~6] Funnel walls (Solid curves, marked by F), centrifugal barriers
(long dashed curves, marked by CF) and shock surfaces (short dashed curves,
marked by S) for the three cases of Fig. 2 for ${\cal E}=1.004$.
In each set, the outermost, middle and the innermost curves are for
$a=0,\ 0.5,\ 0.95$ respectively.

\item[Fig.~7] Entropy function ${\dot{\cal M}}$ is plotted against the
inclination angle $\theta$ for the three cases as in Fig. 3. 
The dashed curved are the values for flows passing through
the outer sonic surfaces ${\dot{\cal M}_o}$, and the solid
curves are for flows passing through the
the inner sonic surfaces ${\dot {\cal M}_i}$. 
Shocks in accretion are possible only when ${\dot {\cal M}_i} >
{\dot{\cal M}_o}$, namely when $\theta_{min,a}<\theta<\theta_{max}=\pi/2$
(equatorial plane). $\theta_{min,a}$ is $\sim 79^o$, $\sim 72^o$
and $70^o$ respectively for $a=0,\ 0.5, \ 0.95$ respectively.
For $\sim 60^o \lsim  \theta<\theta_{min,a}$ shocks may form in winds.

\end{description}

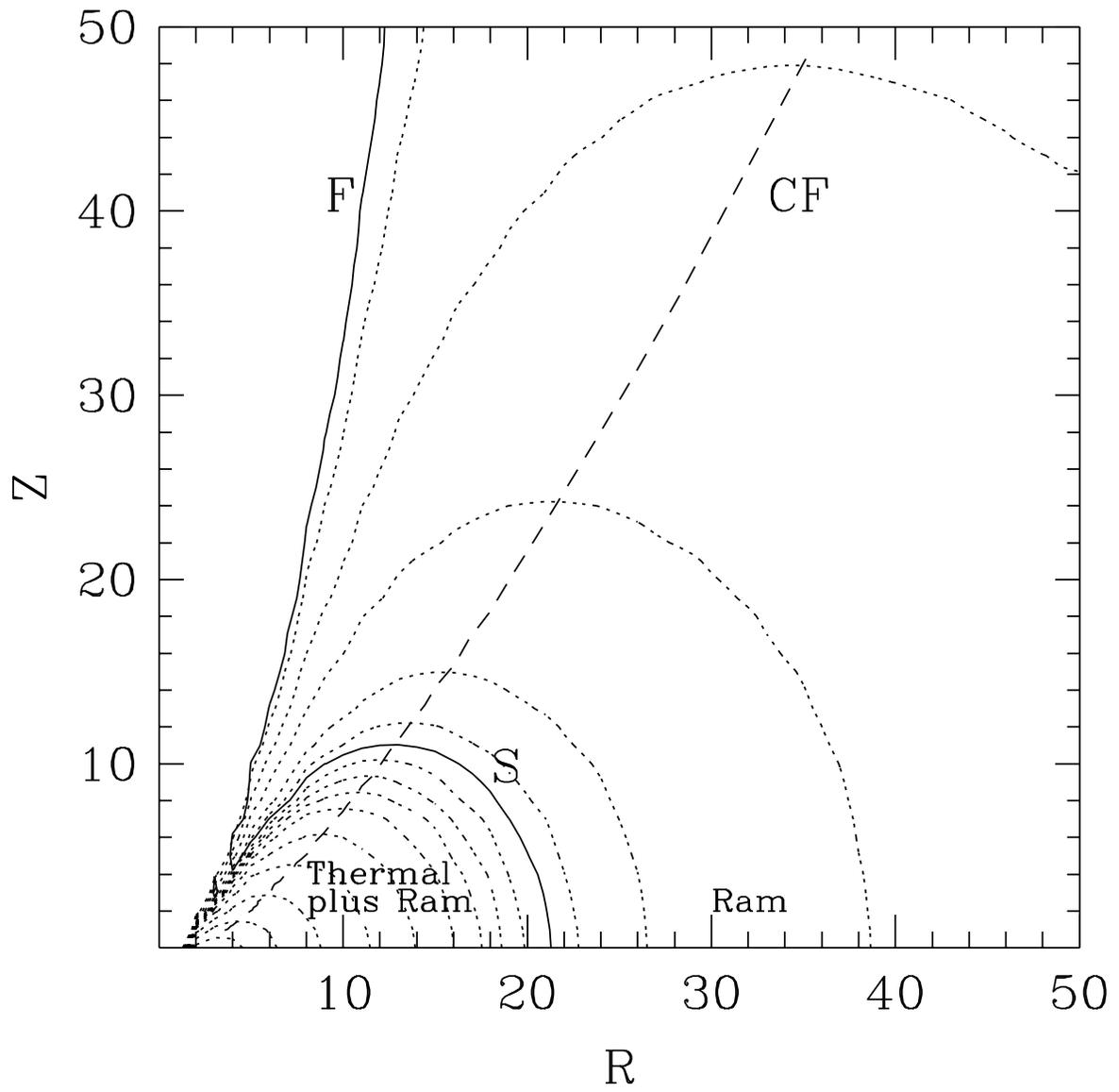
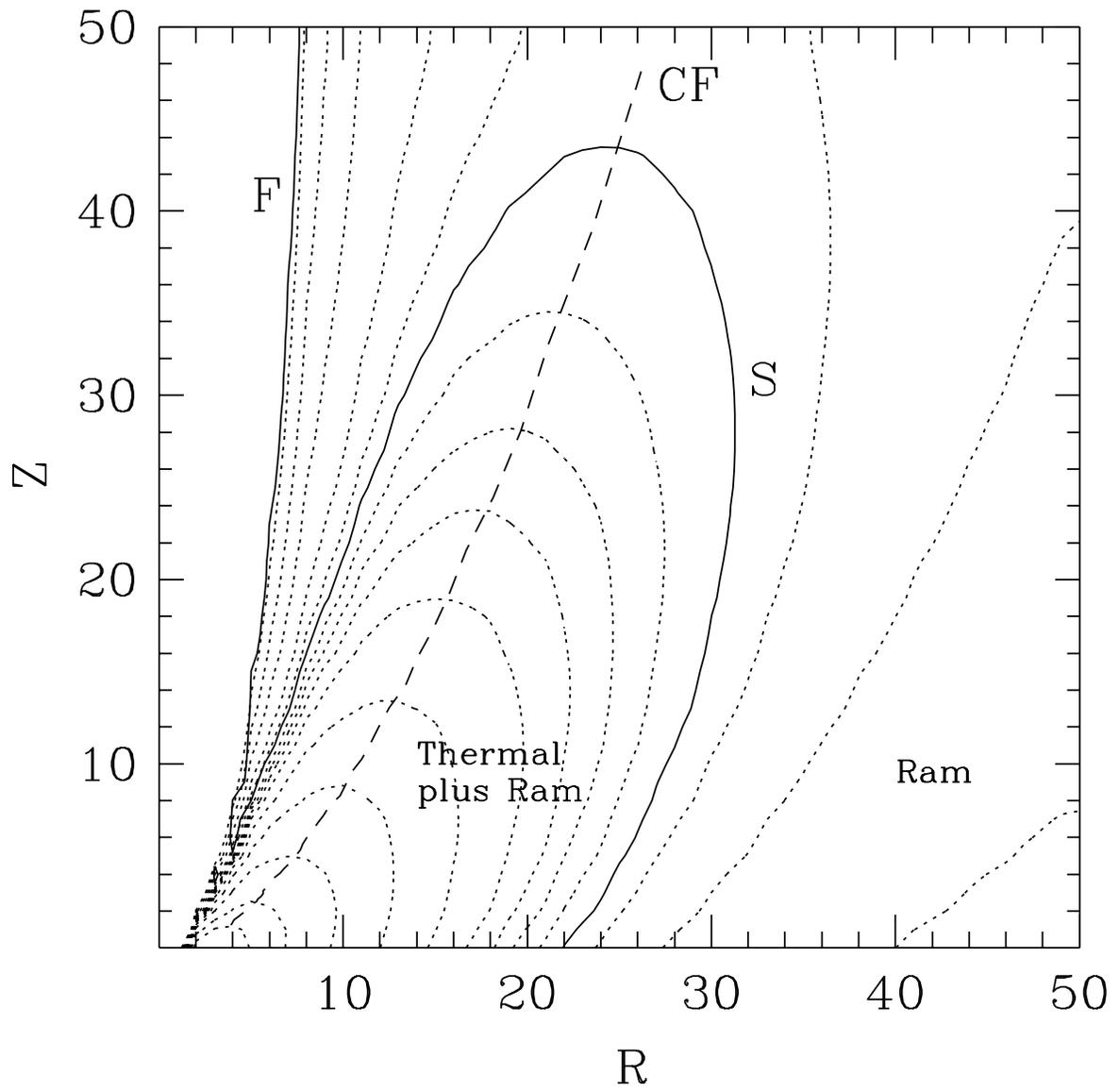
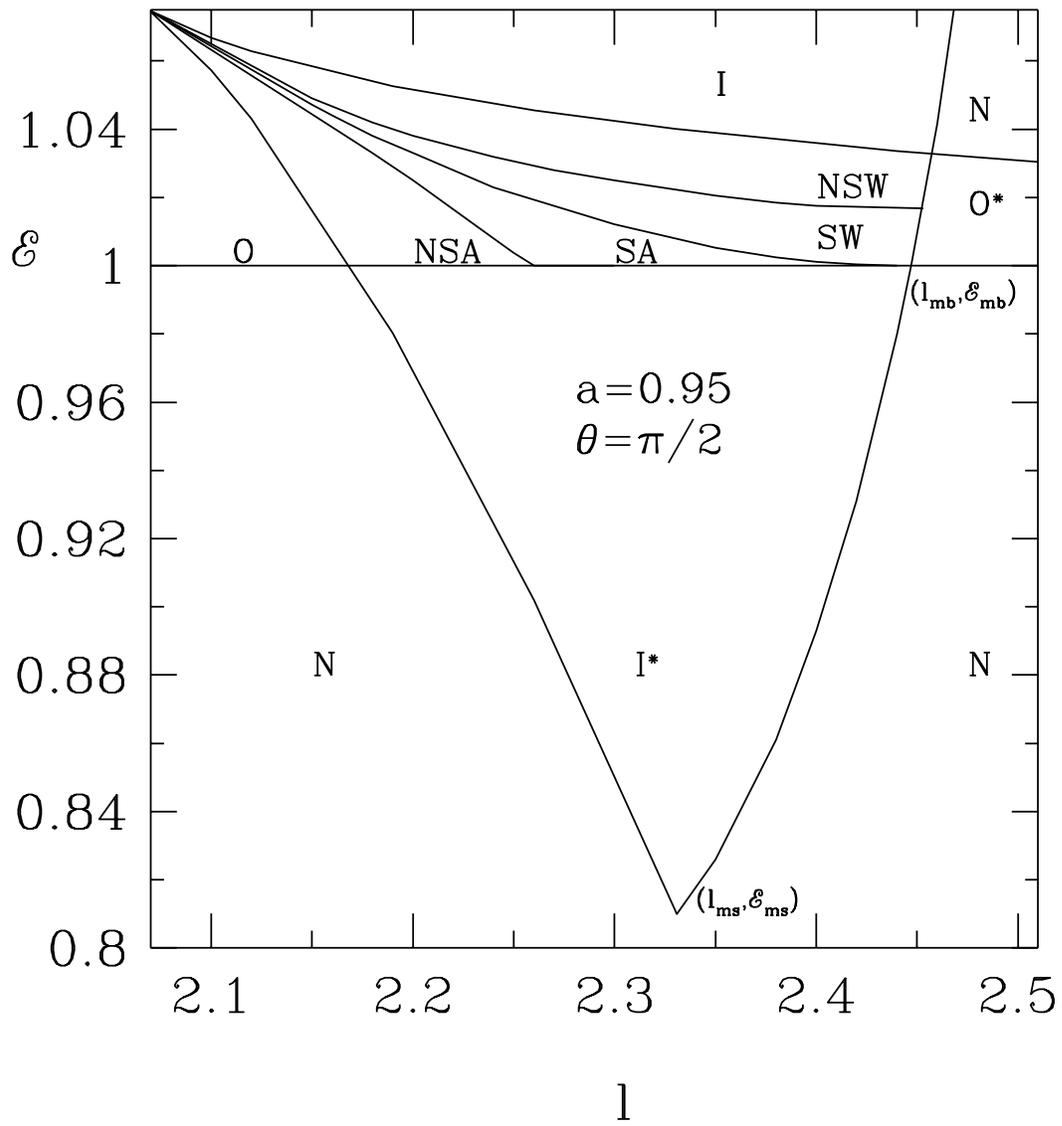
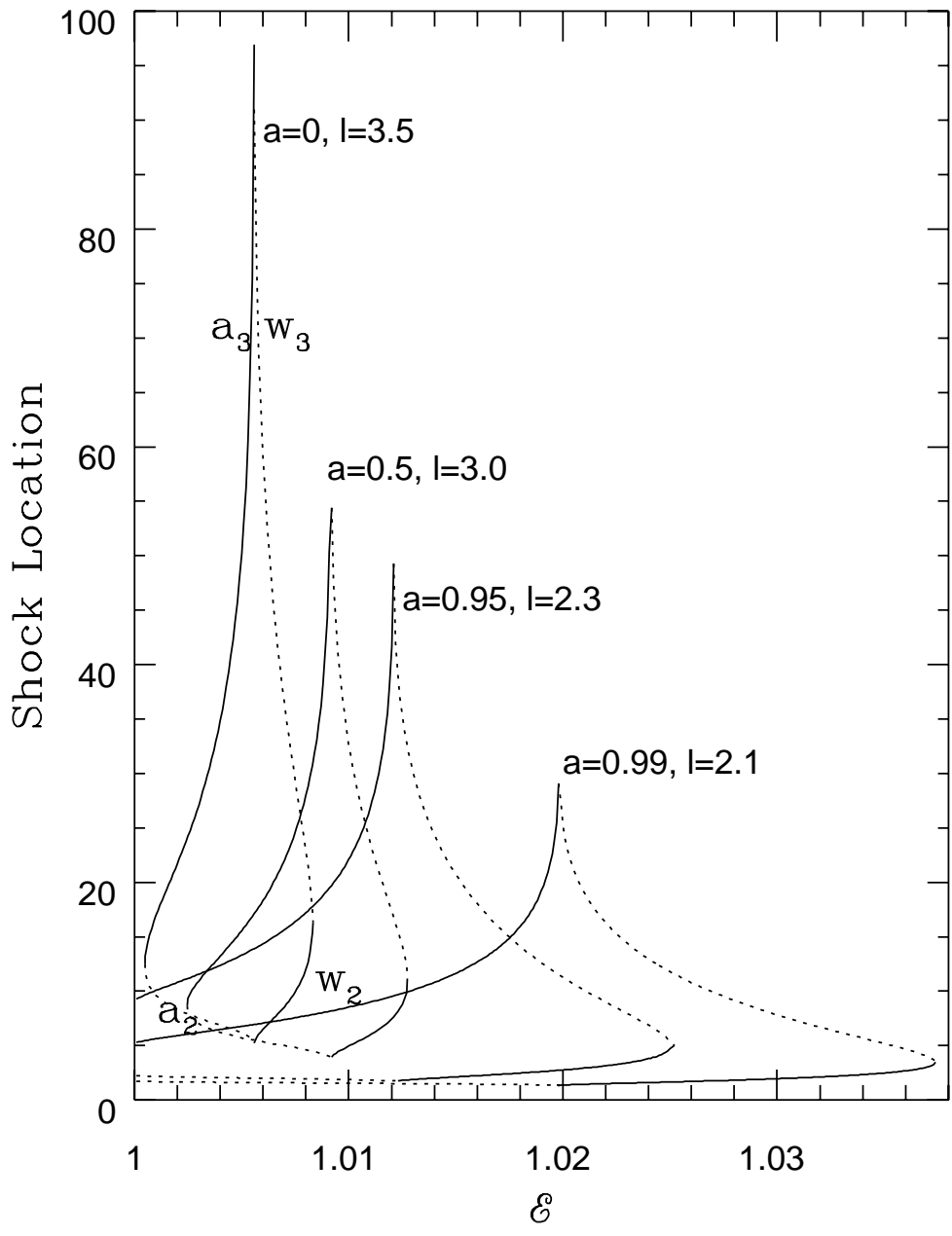
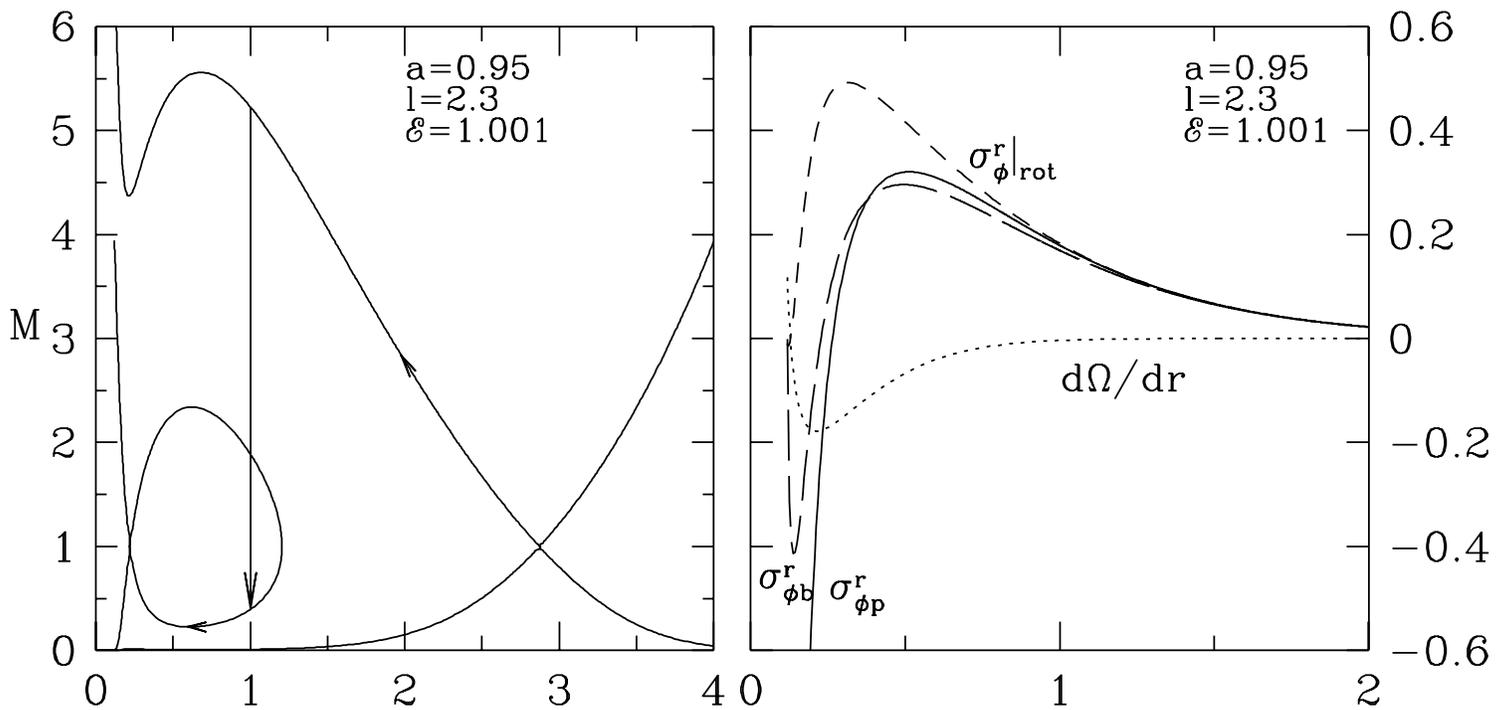
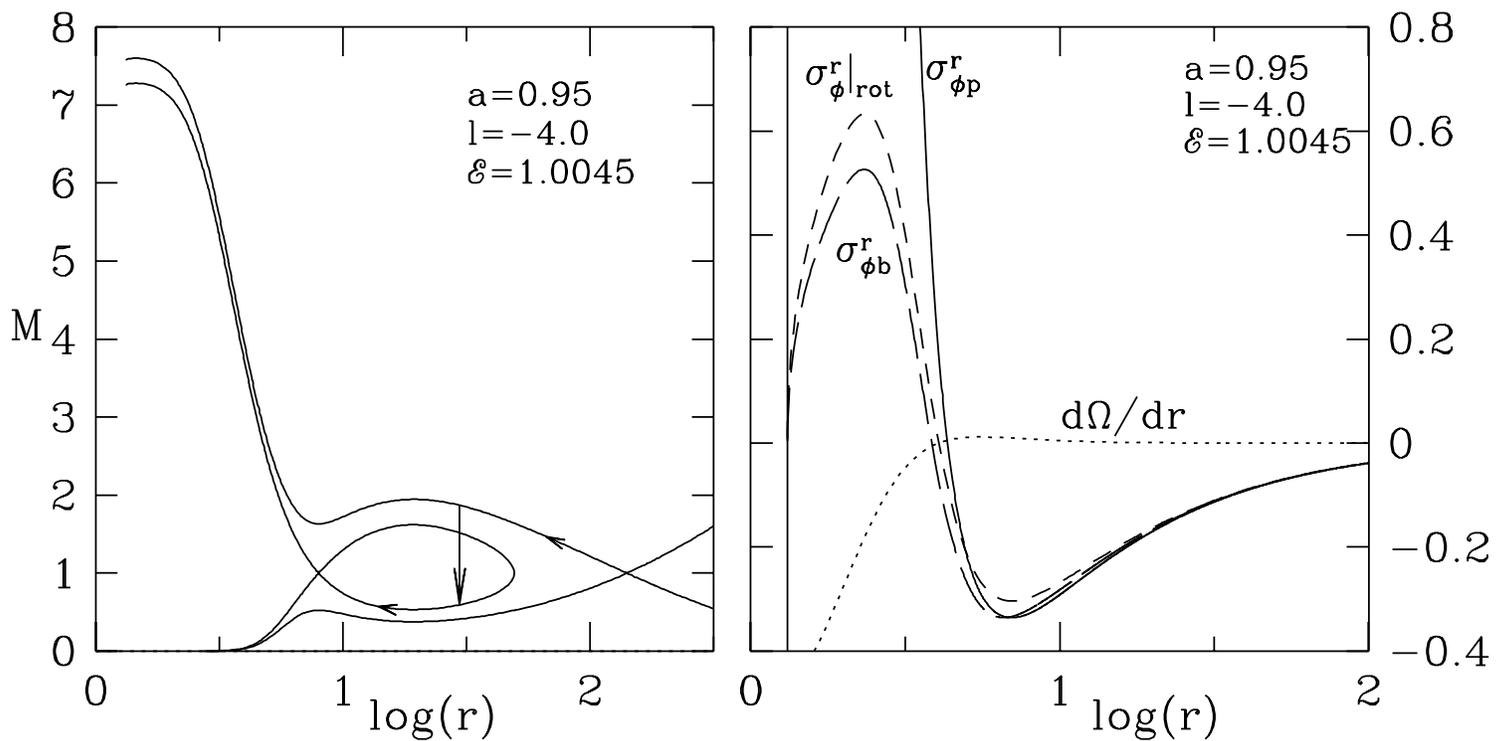
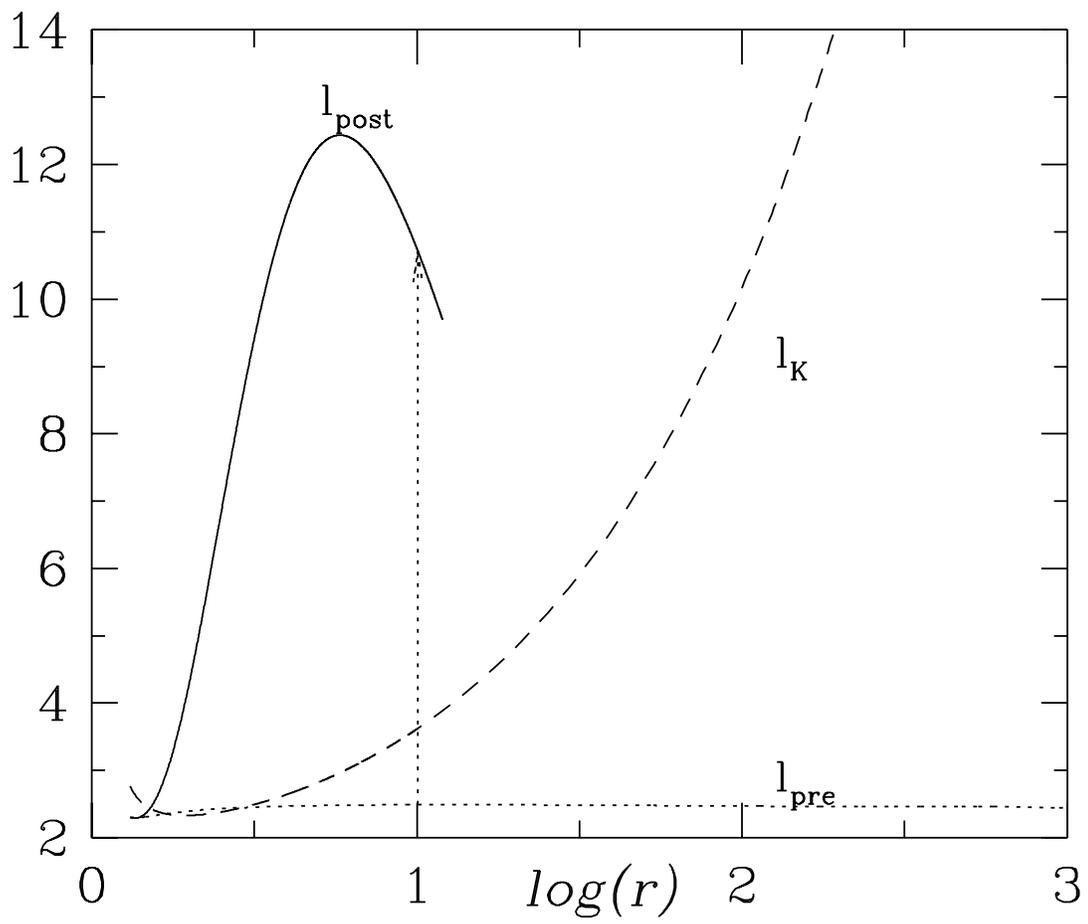
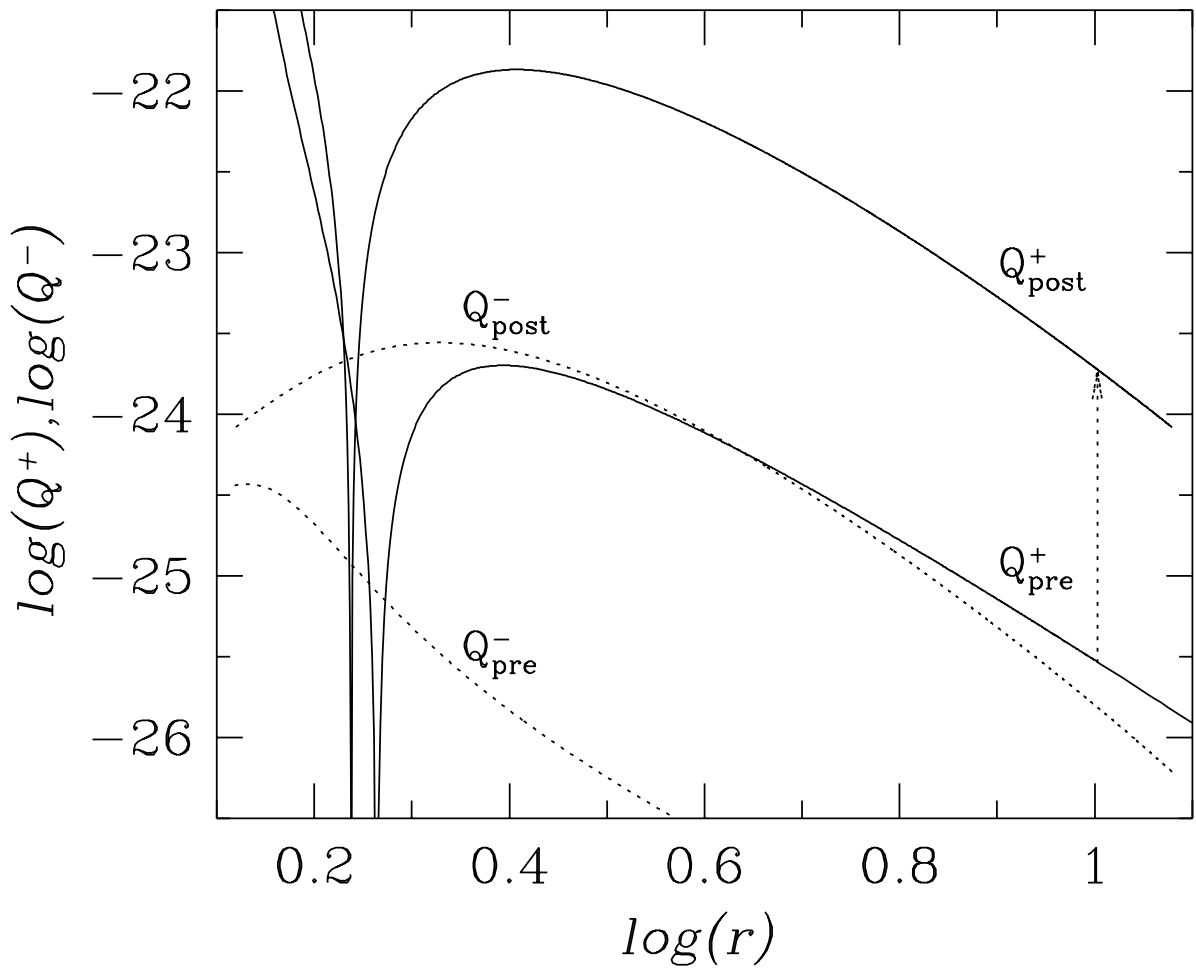
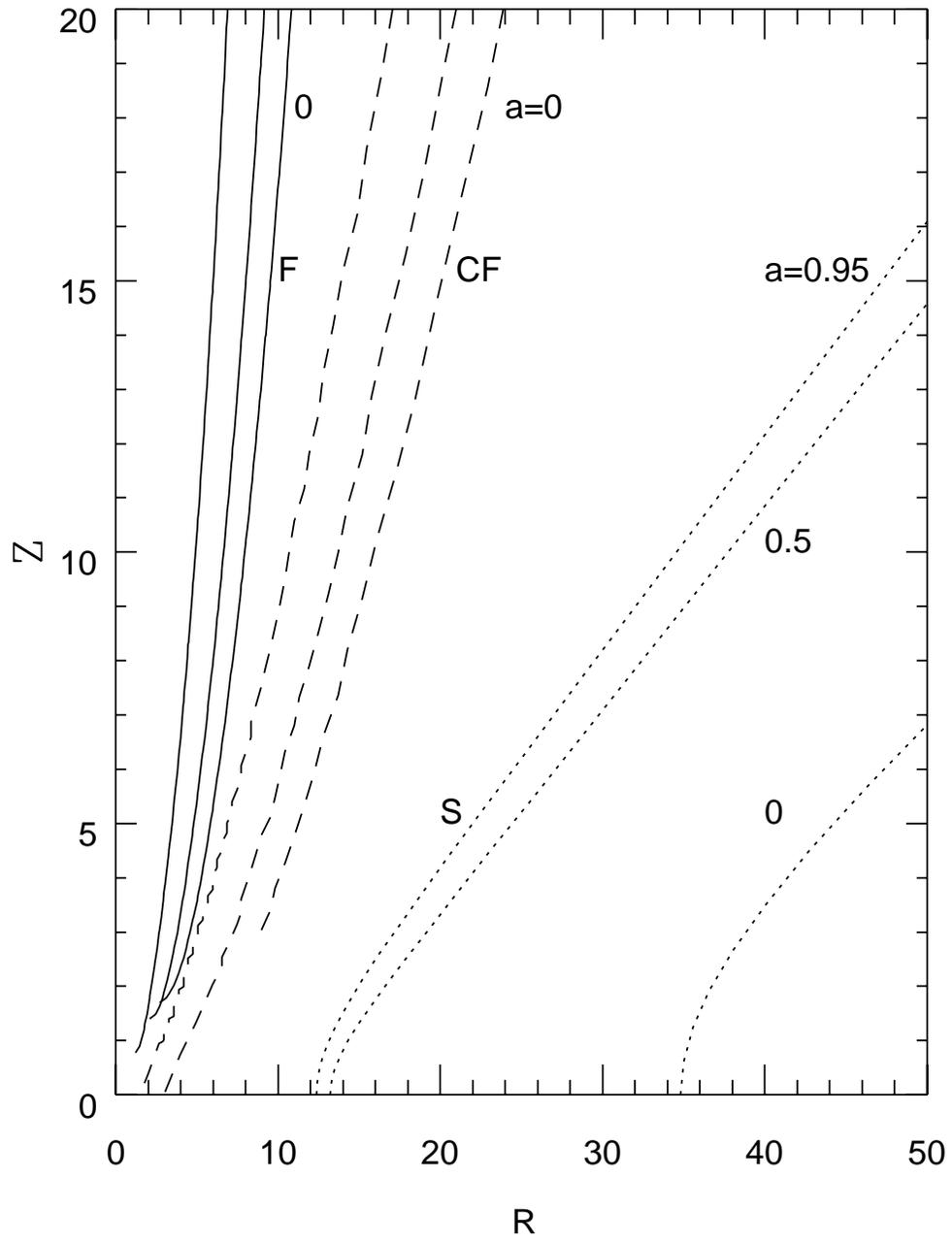
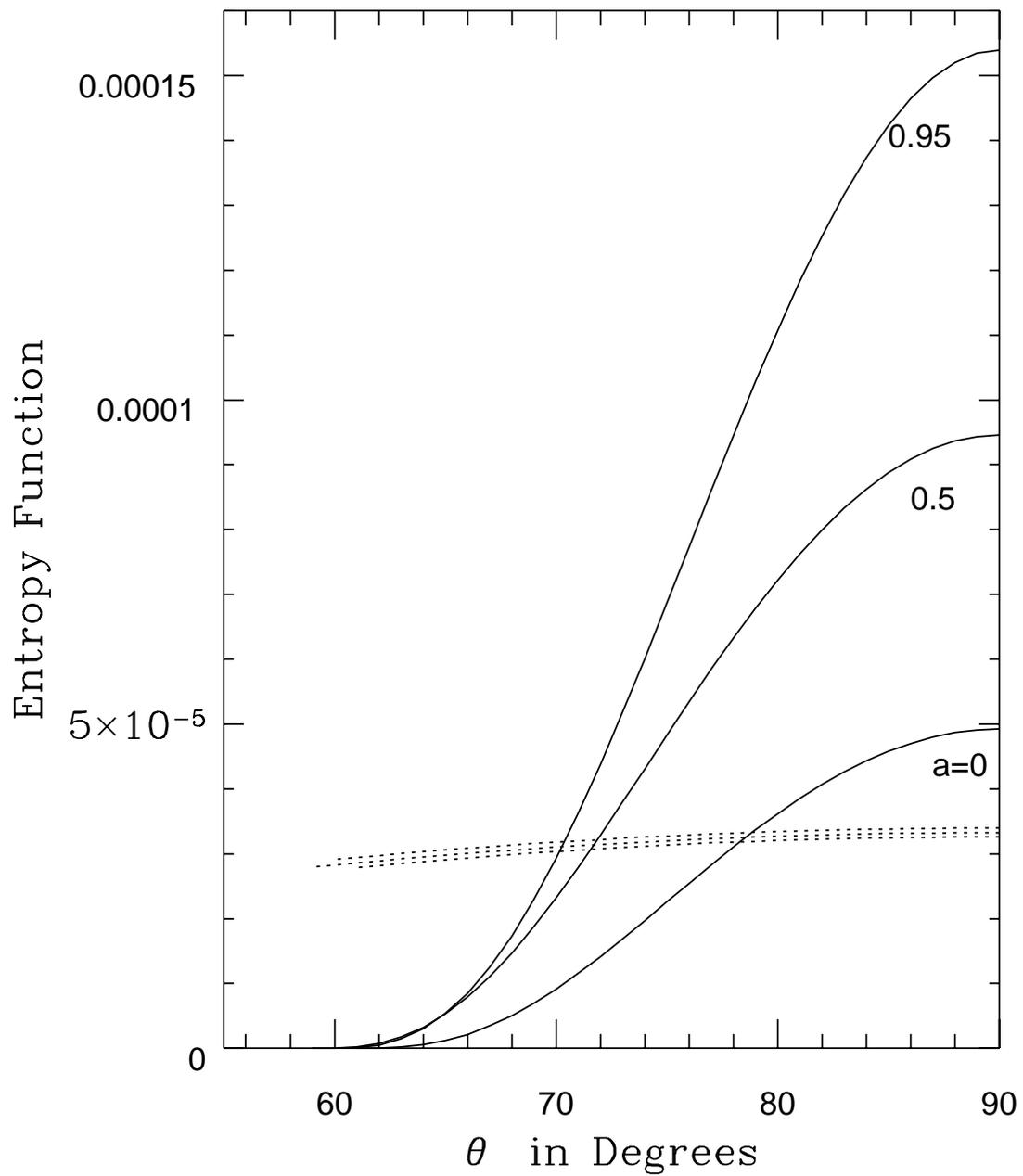
\end{document}